\documentclass{iau}
\usepackage{graphicx}

\title[Spin-orbit resonance of eccentric coorbital bodies] %% give here short title %%
{Spin-orbit coupling and chaotic rotation for eccentric coorbital bodies}

\author[Adrien Leleu, Philippe Robutel \& Alexandre C.M. Correia]   %% give here short author list %%
{Adrien Leleu$^1$,
   Philippe Robutel$^1$ 
   \and 
   Alexandre C.M. Correia$^{2,1}$}

\affiliation{$^1$IMCCE, Observatoire de Paris, CNRS, UPMC Univ. Paris 06, Univ. Lille 1, 77Av.Denfert-Rochereau, 75014 Paris, France \\ email: {\tt aleleu@imcce.fr, robutel@imcce.fr} \\[\affilskip]
$^2$Departamento de F\'isica, I3N, Universidade de Aveiro, Campus de Santiago, 3810-193 Aveiro, Portugal \\email: {\tt correia@ua.pt}}

\pubyear{2014}
\volume{310}  %% insert here IAU Symposium No.
\pagerange{xxx--xxx}
\jname{Complex Planetary Systems}
\editors{Z. Knezevic \& A. Lema\^itre}
\begin{document}

\maketitle

\begin{abstract} The presence of a co-orbital companion induces the splitting of the well known Keplerian spin-orbit resonances. It leads to chaotic rotation when those resonances overlap.

\end{abstract}

\firstsection % if your document starts with a section,

\section{Introduction and Notations}

Given an asymmetric body on a circular orbit, denoting $\theta$ its rotation angle in the plane with respect to the inertial frame, the only possible spin-orbit resonance is the synchronous one $\dot{\theta}=n$, $n$ being the mean motion of the orbit. On an Keplerian eccentric orbit, Wisdom \textit{et al.} (1984) showed that there is a whole family of spin-orbit eccentric resonances, the main ones being $\dot{\theta} = pn/2$ where $p$ is an integer. In 2013, Correia and Robutel showed that in the circular case, the presence of a coorbital companion induced a splitting of the synchronous resonance, forming a family of co-orbital spin-orbit resonances of the form $\dot{\theta} = n \pm k \nu/2$, $\nu$ being the libration frequency in the coorbital resonance. Inside this resonance, the difference of the mean anomaly of the two coorbitals, denoted by $\zeta$, librates around a value close to $\pm \pi/3$ (around the L4 or L5 Lagrangian equilibrium - tadpole configuration), around $\pi$ (encompassing L3, L4 and L5 - horseshoe configuration) or $0$ (quasi-satellite) configuration. We generalize the results of Correia and Robutel (2013) from the case of circular co-orbital orbits to eccentric ones.

\section{Rotation}

 The rotation angle $\theta$ satisfies the differential equation:
\begin{equation}
\ddot{\theta} + \frac{\sigma^2}{2}\left(\frac{a}{r}\right)^3  \sin{2(\theta - f)} =0 , \,\, {\rm with} \,\, \sigma =  n\sqrt{\frac{3(B-A)}{C}} \, ,
\label{eq:rot_gene}
\end{equation}

where $A<B<C$ are the internal momenta of the body, $(r,f)$ the polar coordinates of the center of the studied body and $a$ its instantaneous semi-major axis. 

Let us consider that the orbit is quasi-periodic. As a consequence, the elliptic elements of the body can be expended in Fourier series whose frequencies are the fundamental frequencies of the planetary system. In other words  the time-dependent quantity $ \left( \frac{a}{r} \right)^3 e^{i 2 f}$  that appears in equation (\ref{eq:rot_gene}) reads:

\begin{equation}
  \left( \frac{a}{r} \right)^3 e^{i 2 f} = \sum_{j \geq 0 }\ \rho_j\ e^{(i\eta_jt+ \phi_j )}.
\label{eq:asreqp}
\end{equation}

Where $\eta_j$ are linear combinations with integer coefficients of the fundamental frequencies of the orbital motion (here $n$ and $\nu$) and $\phi_j$ their phases. Thus (\ref{eq:rot_gene}) becomes:

\begin{equation}
\label{eq:theppqp}
\ddot{\theta} = - \frac{\sigma^2}{2} \sum_{j \geq 0 }\ \rho_j \sin\ (2\theta + \eta_j t+ \phi_j).
\end{equation}

For a Keplerian circular orbit, the only spin orbit resonance possible is the synchronous one, since $\rho_0=1$, $\eta_0=2n$, and $\rho_j=\eta_j=0$ for $j>0$. In the general Keplerian case we have the spin-orbit eccentric resonances, $\eta_j = pn$ and the $\rho_j$ are the Hansen coefficients $X^{-3,2}_p(e)$ (see Wisdom \textit{et al.}). For the circular coorbital case, Correia and Robutel (2013) showed that a whole family results from the splitting of the synchronous resonance of the form $\eta_j = 2n \pm k\nu$. For small amplitudes of libration around L4 or L5 (tadpole), the width of the resonant island decreases as $k$ increases.\\ 

\begin{figure}[h]
\begin{center}
 \includegraphics[width=13.5cm]{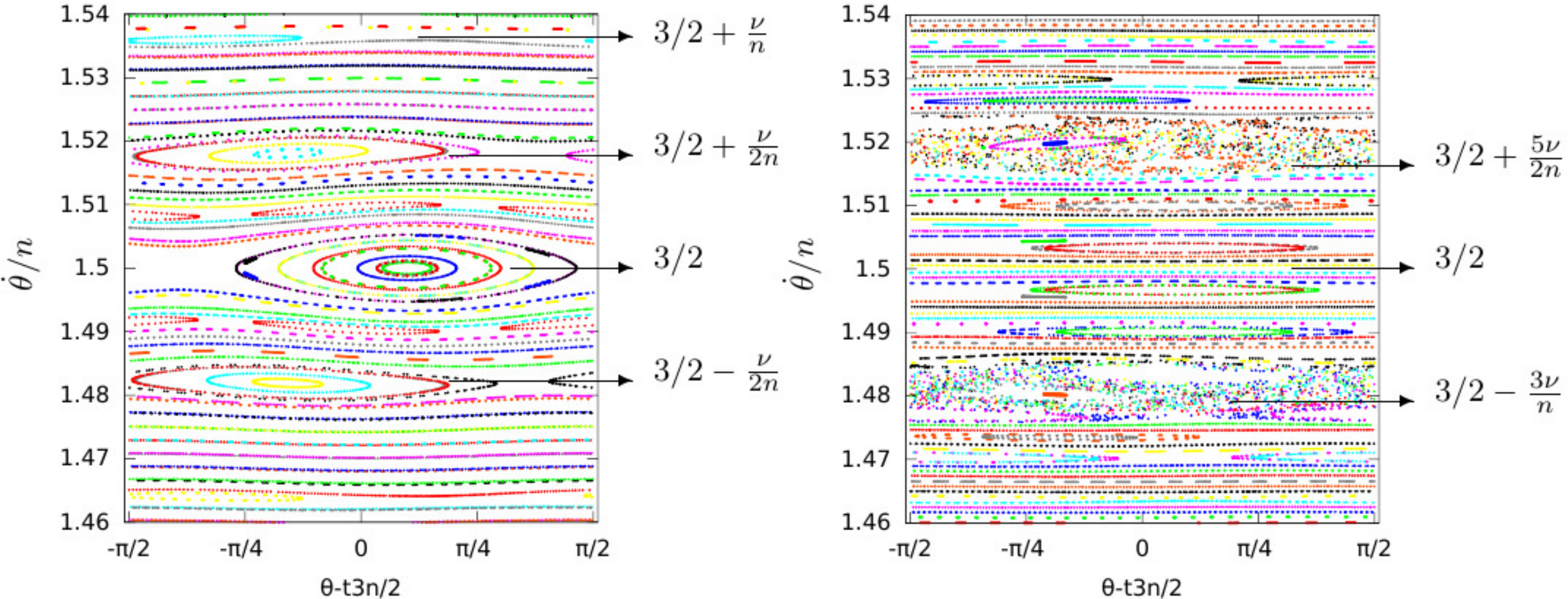} 
 \caption{Poincar\'e surface of section in the plane $(\theta-t\frac{3n}{2},\dot\theta/n)$ near the $3/2$ spin-orbit eccentric resonance. (left): $\zeta_{max}-\zeta_{min}=35^\circ$ - tadpole configuration. (right): $\zeta_{max}-\zeta_{min}=336^\circ$ horseshoe configuration.
 \label{fig}  }
\end{center}
\end{figure}

In the eccentric coorbital case, each eccentric spin-orbit resonance of the Keplerian case splits in resonant multiplets which are centred in $\dot{\theta} =pn/2 \pm k\nu/2$. For relatively low amplitude of libration of $\zeta$, the width of the resonant island decreases as $k$ increases, see Figure \ref{fig} (left). But for higher amplitude, especially for horseshoe orbit, the main resonant island may not be located at $k=0$. In Figure \ref{fig} (right), the main islands are located at $\dot{\theta}=3n/2 \pm 5\nu/2$ and $\dot{\theta}=3n/2 \pm 6\nu/2$. These islands overlap, giving rise to chaotic motion for the spin, while the island located at $\dot{\theta}=3n/2$ is much thinner.

\section{Conclusion}

The coorbital spin-orbit resonances populate the phase space between the eccentric resonances. Generalised chaotic rotation can be achieved when harmonics of co-orbital spin-orbit resonances overlap each other, which is a different mechanism than the one described by Wisdom \textit{et al.} (1984), where the eccentricity harmonics overlap.

\end{document}